\documentclass[showpacs,prb,twocolumn,floats,superscriptaddress]{revtex4}

\usepackage{color}
\usepackage{amsmath,amssymb}
\usepackage{pifont}%
\usepackage{amssymb}  
\usepackage{amsmath}
\usepackage{xcolor}

\usepackage[makeroom]{cancel}
\usepackage{bbold}
\usepackage{float}
\usepackage{tikz}
\usepackage{ulem}
\usetikzlibrary{positioning}
\tikzset{mynode/.style={draw,text width=1.90cm,align=center}
}
\usepackage{makecell}
\usepackage{subfigure}
\usepackage{pifont}   
\usepackage{graphicx} 
\graphicspath{{Figures/}}
\usepackage{dcolumn}  
\usepackage{bm}       
\usepackage{multirow} 
\usepackage{placeins}
\usepackage[colorlinks]{hyperref}

\begin{document}

\title{Effect of parallel magnetic field on Klein tunneling in $pn$ and $pnp$ graphene\\ trilayer junctions}
\date{\today}
\author{Abderrahim El Mouhafid}
\email{elmouhafid.a@ucd.ac.ma}
\affiliation{Laboratory of Theoretical Physics, Faculty of Sciences, Choua\"ib Doukkali University, PO Box 20, 24000 El Jadida, Morocco}
\author{Ahmed Jellal}
\email{a.jellal@ucd.ac.ma}
\affiliation{Laboratory of Theoretical Physics, Faculty of Sciences, Choua\"ib Doukkali University, PO Box 20, 24000 El Jadida, Morocco}
\affiliation{Canadian Quantum Research Center, 204-3002 32 Ave Vernon, \\ BC V1T 2L7, Canada}

\begin{abstract}
        

The effect of a parallel magnetic field $B$ and a potential substrate on the transmission, conductivity, and Fano factor of a biased ABC-stacked $pn$ and $pnp$ trilayer graphene junction (ABC-TLG) is investigated theoretically at low energy. We discovered that, in the presence of a high magnetic field $B=1400$T, the ABC-TLG can exhibit new Klein tunneling at two new incidence $k_y$ values, in addition to the usual value of $k_y=0$ and whatever the energy value is. Indeed, the transmission of ABC-TLG via $ pn $ and $ pnp $ junctions is transformed into a three-separate transmission of single-like graphene (SLG), with Klein tunneling appearing at $k_y=0$ and $k_y=\pm\kappa$ ($\kappa$ is highly dependent on $B$). Furthermore, the conductivity $\sigma$ and Fano factor $F$ behave similarly to SLG, with ABC-TLG having the lowest $\sigma=3g_0/\pi$ and highest $F=1/3$ at $B=1400$T and $\sigma_{\text{TLG}}=3\sigma_{\text{SLG}}$. 

\end{abstract}

\maketitle
\section{Introduction}
Graphene, the two-dimensional allotrope of carbon, has drawn an enormous amount of attention in the literature after its first isolation on an oxide substrate\cite{Novoselov306,Novoselov197}. Apart from the theoretical physics point of view\cite{Geim183,Novoselov177}, graphene has emerged as a possible candidate for different electronic devices, including field effect transistors\cite{Lemme282,Oostinga151,Chen206,Lin262}. However, the small band gap of graphene reduces the controllability of such devices and thus limits their widespread applications. On the other hand, a three-stacked monolayer graphene (Rhombohedral stacking) (ABC-TLG)\cite{Avetisyan115432,Morpurgo625} has been shown to provide a significant band gap\cite{Quhe1794,Aoki123,Castro109,Latil036803,Kumar163102,vanduppen226101,Kumar222101,vanduppen195439,Craciun383,Avetisyan115432,Tang9458,Lui944,Bao948,Jellal534} that is considered as a potential channel material in field effect transistors\cite{Umoh805,Sadeghi1250047,Dongwei105303,Rahmani55}. 
More recently, it was found that such an electric field causes an energy gap in ABC-TLG, which was found to be a non-monotonic function of the gate voltage, and a re-entrant opening and closing of the gap was predicted as a function of the electric field strength\cite{Quhe1794,Aoki123,Castro109,Latil036803,Kumar163102,vanduppen226101,Kumar222101,vanduppen195439,Craciun383,Avetisyan115432,Tang9458,Lui944,Bao948,Jellal534,Avetisyan035421,Uddin56}. 
The electronic band structure of the ABC-TLG in the presence of back and top gates was investigated in Ref.\cite{Zou369,Salah203704} and the transport properties were investigated in Ref.\cite{Kumar163102,vanduppen226101,Kumar222101,vanduppen195439,Jellal534}.

%
        
        In this paper, we analyze the effects of a parallel magnetic field on the quantum transport in a biased ABC-TLG, see Fig. \ref{Schematicsystem}(b). In a high magnetic field, the transmission probability is equivalent to the three-separate transmission of single-like graphene (SLG). Furthermore, the minimum conductivity $g_0/\pi$ and corresponding Fano factor $1/3$ found in SLG are found in ABC-TLG such as $\sigma_{\text{TLG}}=3\sigma_{\text{SLG}}$ at $B=1400$T. Our results may help to understand the electronic properties of the ABC-TLG in the presence of a strong parallel magnetic field that cannot be created in terrestrial laboratories.
        
%

The paper is organized as follows. In Sec. \ref{ELECTRONIC MODEL} we investigate the electronic properties of the system. We analyze the electronic band structure of ABC-TLG for pristine TLG and with parallel magnetic field, inter-layer bias, and both magnetic field and bias. We present in detail the formalism used for the calculation of the transmission with the corresponding conductivity and Fano factor for our system. In Sec. \ref{numerical results}, we  numerically analyze our  results and give different discussions. Our conclusions are presented in Sec. \ref{conclusions}.
\section{ELECTRONIC MODEL}
\label{ELECTRONIC MODEL}
\begin{figure}[t!]
\vspace{0.cm}
\centering\graphicspath{{./Figures/}}
\includegraphics[width=\linewidth]{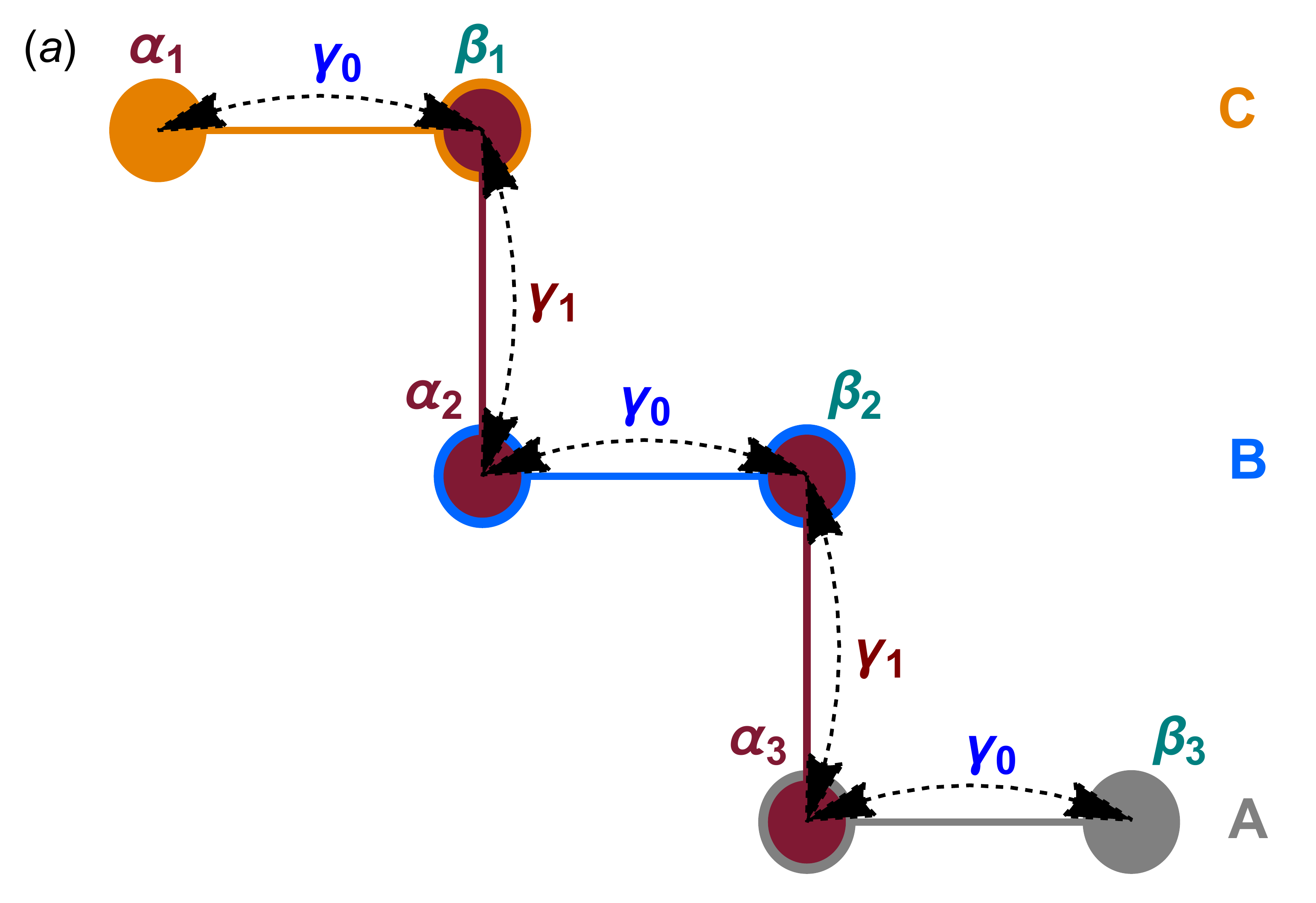}\\
\includegraphics[width=\linewidth]{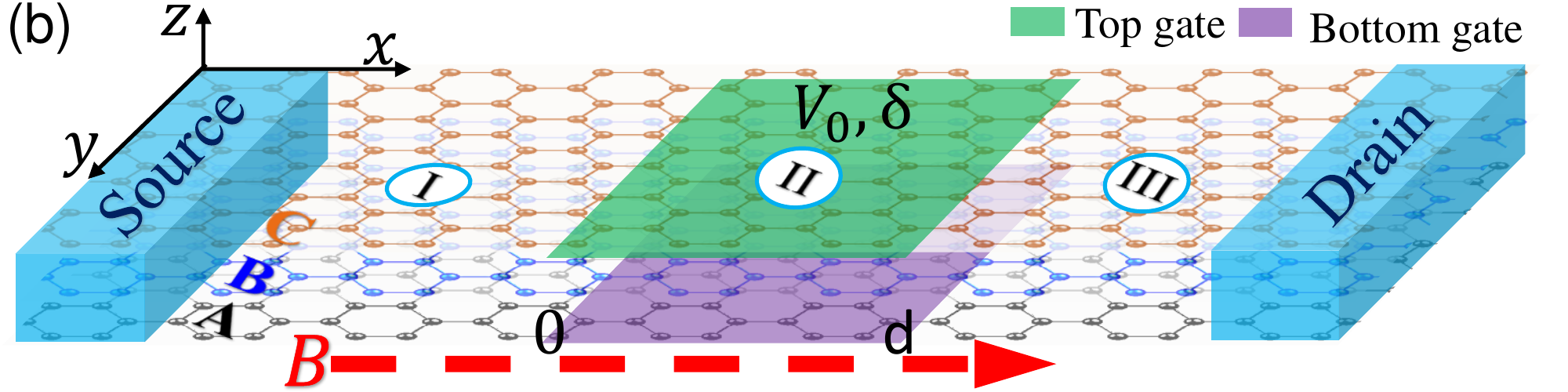}\\
\includegraphics[width=\linewidth]{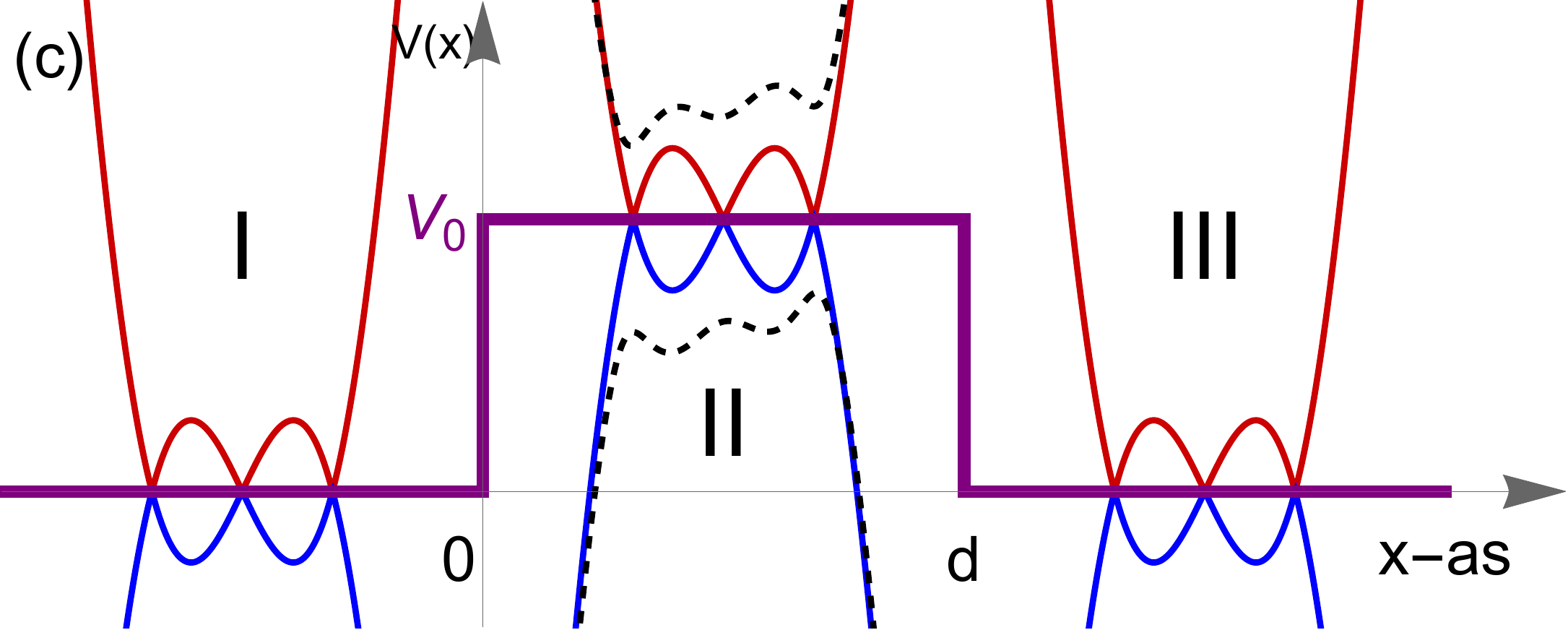}\\
\caption{(Color online) (a) Lattice structure of ABC-stacked graphene trilayer A, B, and C are three distinct positions of the hexagonal lattice when graphene trilayers are stacked. (b) Schematic pictures of the proposed system with the parameters of a rectangular barrier configuration. There is an electrostatic potential difference $\delta$ between the top and bottom layers applied in region II. The electrostatic potential $V_0$ is the same for all three layers applied in the previous region. The parallel magnetic field that is considered in this work is indicated by red arrows on the bottom-hand side of this figure. (c) Profile of a rectangular barrier applied to our system as well as the corresponding band regions. The dashed and thick curves represent the band for biased and unbiased  ABC-TLG in the presence of a parallel magnetic field, respectively.}\label{Schematicsystem}
\end{figure}

We consider an ABC-TLG lattice with three coupled layers\cite{Koshino165409, Koshino125304} separated by $c_\perp=0.333$ nm\cite{Celal012022} and connected by $\gamma_1=0.377$ eV\cite{Koshino165409,Koshino125304}, each with carbon atoms arranged on a honeycomb lattice designed by $A$, $B$, $C$, including pairs of inequivalent sites $\{\alpha_{1},\beta_{1}\}$, $\{\alpha_{2},\beta_{2}\}$ and $\{\alpha_{3},\beta_{3}\}$ in the top, center, and bottom layers, respectively, as shown in Fig. \ref{Schematicsystem}(a).
Each layer has two sublattices, $\alpha_j$ and $\beta_j$ ($j$ denotes the layer index), with an inter-atomic distance of $a_0=0.142$ nm and an intra-layer coupling of $\gamma_0=3.16$ eV\cite{Koshino165409}. Based on the profile of a rectangular barrier depicted in Figs. \ref{Schematicsystem}(b,c), we set all regions composing our system as $j=\text{I}$ ($x\leq 0$), $j=\text{II}$ ($0<x\leq d$), and $j=\text{III}$ ($x>c$). 

%
The continuum Hamiltonian written in the basis $\vec\Psi=\left(\vec\Psi_1,\vec\Psi_2 ,\vec\Psi_3\right)^T$, with $\vec\Psi_i=\left( \phi_{A_i},\phi_{B_i},\phi_{C_i} \right)$ and $i=1,2,3$, can describe the carrier dynamics at low energy in the $j$-th region. This is\cite{vanduppen195439,Jellal534}
\begin{eqnarray}
\mathcal{H}(\vec p)=\left[
\begin{array}{ccccc}
\mathcal{H}_1(\vec p) & \Gamma & 0 \\
\Gamma^{\dag } & \mathcal{H}_2(\vec p) & \Gamma \\
0 & \Gamma^{\dag} & \mathcal{H}_3(\vec p)%
\end{array}%
\right]\label{HamABC},
\end{eqnarray}
where the interlayer coupling $\Gamma$ is  given by
\begin{eqnarray}
\Gamma=\left[
\begin{array}{cc}
0 & 0 \\
\gamma_{1} & 0%
\end{array}%
\right]\label{Gamma2},
\end{eqnarray} 
and $\mathcal{H}_i(\vec p)=v_F \vec\sigma\cdot\vec p+V_{i}(x)I_2$ is the SLG Hamiltonian of the $i$-th layer, $v_F=(\sqrt{3}/2)a_0\gamma_0/\hbar$ is the Fermi velocity, $\vec\sigma=\{\sigma_x,\sigma_y\}$ are the Pauli matrices, $\vec p=\{p_{x},p_{y}\}$ denotes  the in-plane momentum, $V_i(x)$ is the electrostatic potential, and $I_2$ is the $2\times2$ unit matrix. When there is a finite bias, an electric field of magnitude $\left\vert V_3-V_1 \right\vert/c_\perp$  can be induced between the top and bottom layers, i.e., $V_1\neq V_3.$ We can now introduce a magnetic field of the form $\vec B=B (\cos \theta\hat x+\sin \theta\hat y)$ that runs parallel to the layers, where $\theta$ is the angle between the $x$-axis and the direction of $\vec B$. The presence of a parallel magnetic field can be incorporated into Eq. \eqref{HamABC} using the Peierls substitution $\vec p\rightarrow \vec p+e \vec A$, where $\vec A=B z(\sin \theta\hat x-\cos \theta\hat y)$ is a vector potential. When an electron hops between the layers, it will experience a Lorentz force of the form    $\vec F=-e\left( \vec v \times \vec B\ \right)$. 
%
Assume that layer 1 is at $z=-c_\perp/2 $ and layer 3 is at $z=c_\perp/2$, resulting in a total momentum shift of $\Delta\vec p=\int \vec Fdt= eBc_{\perp}\left( \sin \theta\hat x -\cos \theta \hat y\right)$ with   $\left\vert \Delta\vec p/\hbar\right\vert=2\left\vert \vec\kappa \right\vert=c_\perp/l^{2}_{B}=5.1655\times 10^3\ B[T]\ \text{cm}^{-1}$ where $\vec\kappa$ is the shift in the wavevector and $l_{B}=\sqrt{\hbar/eB}$ is the magnetic length.
%
If momentum is $\vec p$ at $z = 0$, momentum will be shifted in layers 1 and 3 as $\vec p=\vec p\pm\hbar \vec\kappa$, respectively. 


Before we get into the energy spectrum, let us define the ratio $a=l/a_0$ in relation to the inter-layer length $l=\hbar v_F/\gamma_1=1.76$ nm. 
This allows us to define the following dimensionless quantities:
\begin{align*}
E\rightarrow\frac{E}{\gamma_1}, V_0\rightarrow\frac{V_0}{\gamma_1}, \delta\rightarrow\frac{\delta}{\gamma_1}, (\vec k,\kappa)\rightarrow a(\vec k,\kappa),\ \vec r\rightarrow\frac{\vec r}{a},
\end{align*}
where $E$ is the incident energy, $V_0$ is the strength of
$pn$ and $pnp$ electrostatic gate junctions along  the $ x $-axis, such as
\begin{align}
&V_{pn}=\begin{cases}0 & \mbox{if } x<0 \\
V_0I_{6}+\Delta & \mbox{if } x\geqslant0, \\
\end{cases}\label{potential}
\\
&
V_{pnp}=\begin{cases}0 & \mbox{if } x<d \\
V_0I_{6}+\Delta & \mbox{if } 0\leqslant x\leqslant d \\
0 & \mbox{if } x>d, \\
\end{cases}\label{potential}
\end{align}
$ I_{6} $ is $6\times 6$ unit matrix, $\Delta=\text{diag}[\delta,\delta,0,0,-\delta,-\delta]$ is a diagonal matrix, and $\delta$ is the interlayer potential difference between the top and bottom layers.

\begin{figure}[t!]
\vspace{0.cm}
\centering\graphicspath{{./Figures/}}
\includegraphics[width=\linewidth]{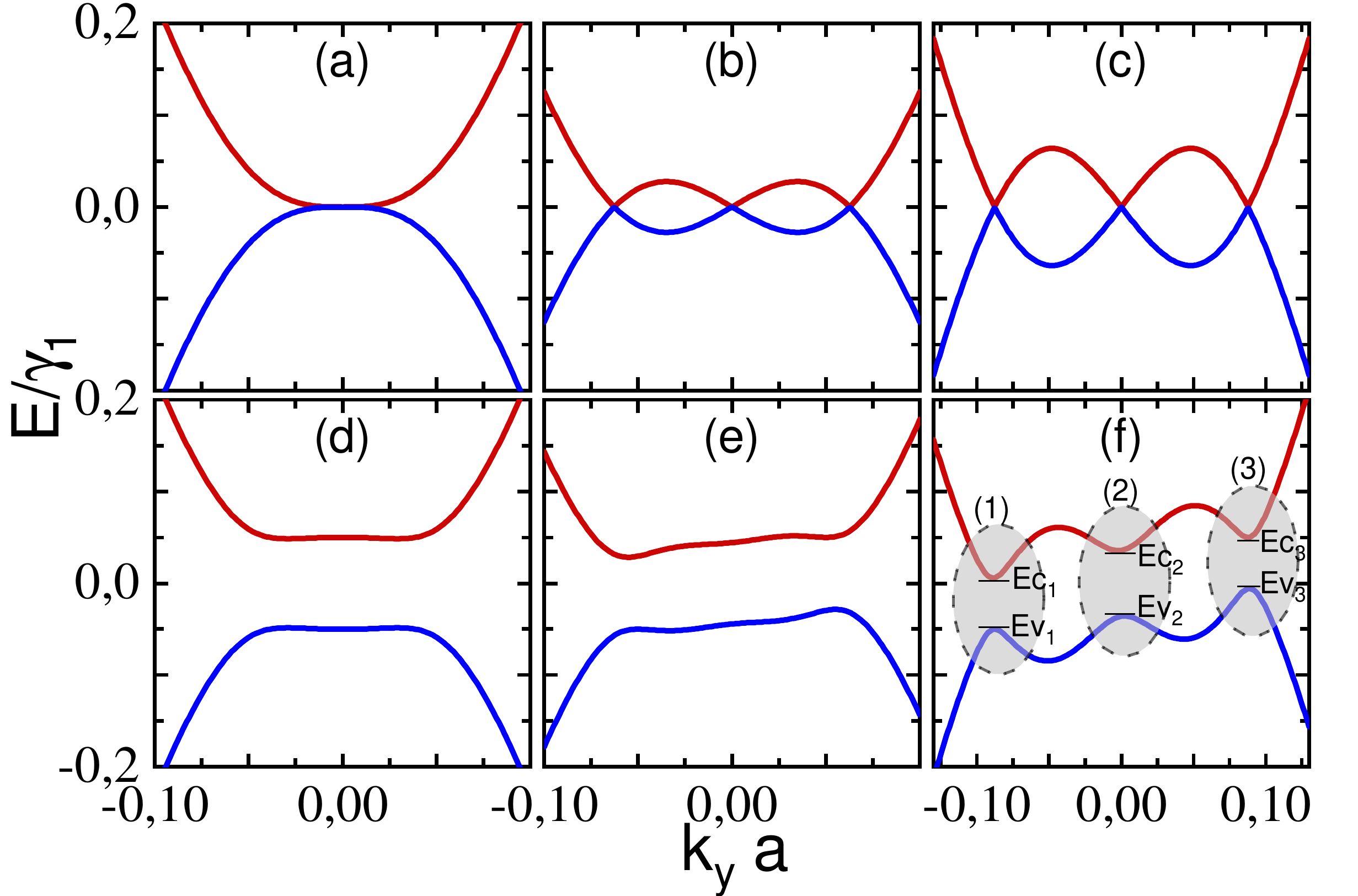}\\
\includegraphics[width=1.2in]{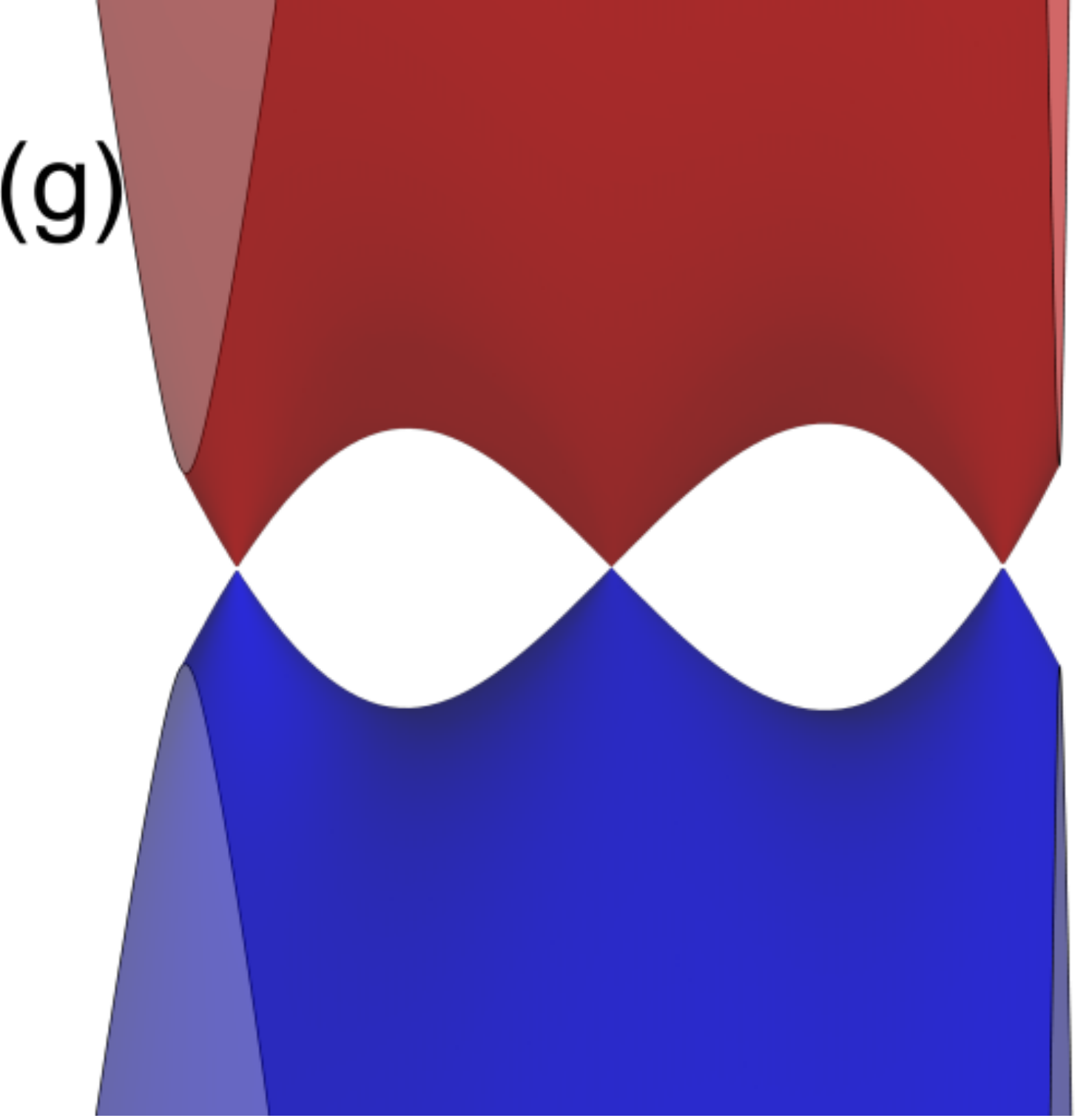}\includegraphics[width=1.2in]{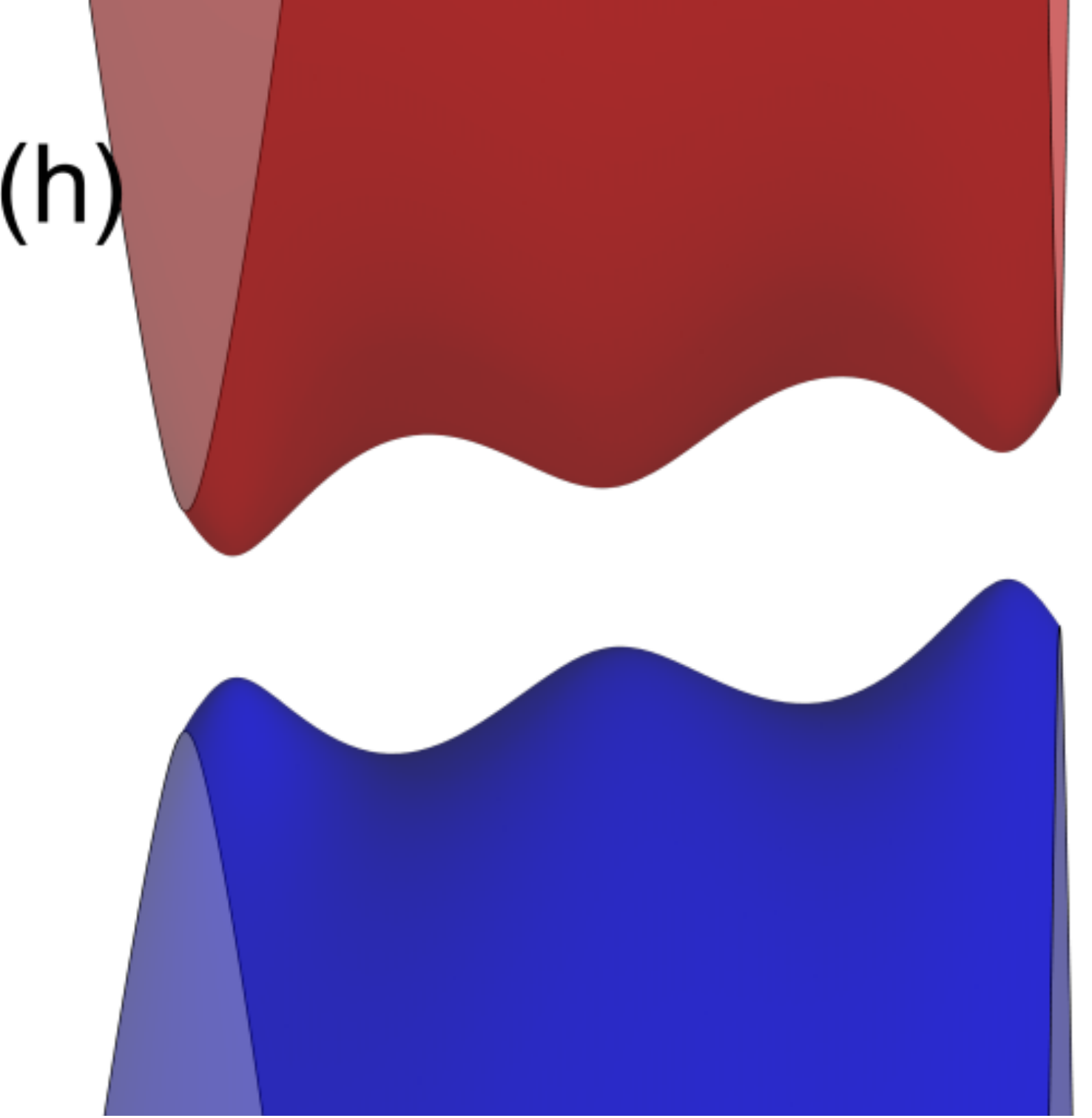}
\caption{(Color online)
ABC-TLG electronic bands (a) pure TLG, (b,c) parallel magnetic field $B$, (d) inter-layer bias $\delta$, and (e,f) both $B$ and $\delta$. The top row has $\delta=0$ and the bottom row has $\delta=0.05\gamma_1$. $B=0$T, $800$T and $1400$T in the left, middle, and right row. $E_{\text{c}_i}$ and $E_{\text{v}_i}$ ($i=1, 2, 3$) are the minimums of the ABC-TLG conduction and valence bands separated by an energy gap, $E\text{g}_i=E_{\text{c}_i}-E_{\text{v}_i}$. As will be shown in Sec. \ref{numerical results}, the three gray circles behave exactly like the energy spectrum of SLG in the absence of $ B $ inside the barrier ($pn$ or $pnp$ junction) of height $V_{0}=V_{0_i}=(E_{\text{c}_i}+E_{\text{v}_i})/2$ with the inter-layer bias $\delta=\delta_i=(E_{\text{c}_i}-E_{\text{v}_i})/2$. $ 3D $ energy spectrum's (g) and (h) associated with cases (c) and (f), respectively.                          
}\label{EnergyBdelta}
\end{figure}

The energy spectrum of the Hamiltonian \ref{HamABC} with and without inter-layer bias in the presence of the parallel magnetic field, oriented along the $x$-direction, is depicted in Fig. \ref{EnergyBdelta}. Pristine TLG has gapless and parabolic bands\cite{Aoki123,Castro109,Latil036803,Kumar163102,vanduppen226101,Kumar222101,vanduppen195439,Jellal534}, see Fig. \ref{EnergyBdelta}(a), while the magnetic filed keeps the spectrum gapless but splits the cone  into  three located at $k_y=\pm\left\vert  \vec\kappa \right\vert$ and at $k_y=0$, as shown in Figs. \ref{EnergyBdelta}(b,c).  A bias $\delta$, on the other hand, opens  a direct energy gap of $2\delta$ in the spectrum\cite{Aoki123,Castro109,Kumar163102,Kumar222101,vanduppen195439},  as shown in Fig.\ \ref{EnergyBdelta}(d). However, in the presence of both $B$ and $\delta$ the spectrum becomes tilted and the gap is now indirect with breaking electron-hole symmetry, see Figs. \ref{EnergyBdelta}(e,f). Note that  influence of parallel magnetic fields is significant only for strong magnetic fields which cannot be created in terrestrial laboratories. However, mechanical deformation like in BLG\cite{Roy2013,Koshino2013,He2014,Daboussi2014} and in TLG \cite{Chen2563,Park249,Haoxin6582,Fischer5,Alexander2608}, such as a twist, shift or strain, can provide strong pseudo-parallel magnetic fields in BLG.  This pseudo magnetic field has the same local effects in the Brillouin zone as the real ones, but the global effect is different\cite{Donck2016a}.  Since we are interested in low energy and around the corner of the Brillouin zone, we can safely adopt this scenario.  For example, if we establish  a mismatch $\lambda$ between the two layers in BLG\cite{Koshino2013} this will produce a pseudo magnetic field and result in a shift in the momentum.   

To calculate the transmission probabilities, the desired solution in each region $j$ must be obtained. Then, implementing the transfer matrix together with appropriate boundary conditions gives the transmission and reflection probabilities\cite{vanduppen195439,Jellal534}. These results will be used to deal with different issues related to our system. Indeed, we will compute transmission and reflection channels together with the associated conductivity and Fano factor.

\begin{figure}[t!]
\vspace{0.cm}
\centering\graphicspath{{./Figures/}}
\includegraphics[width=\linewidth]{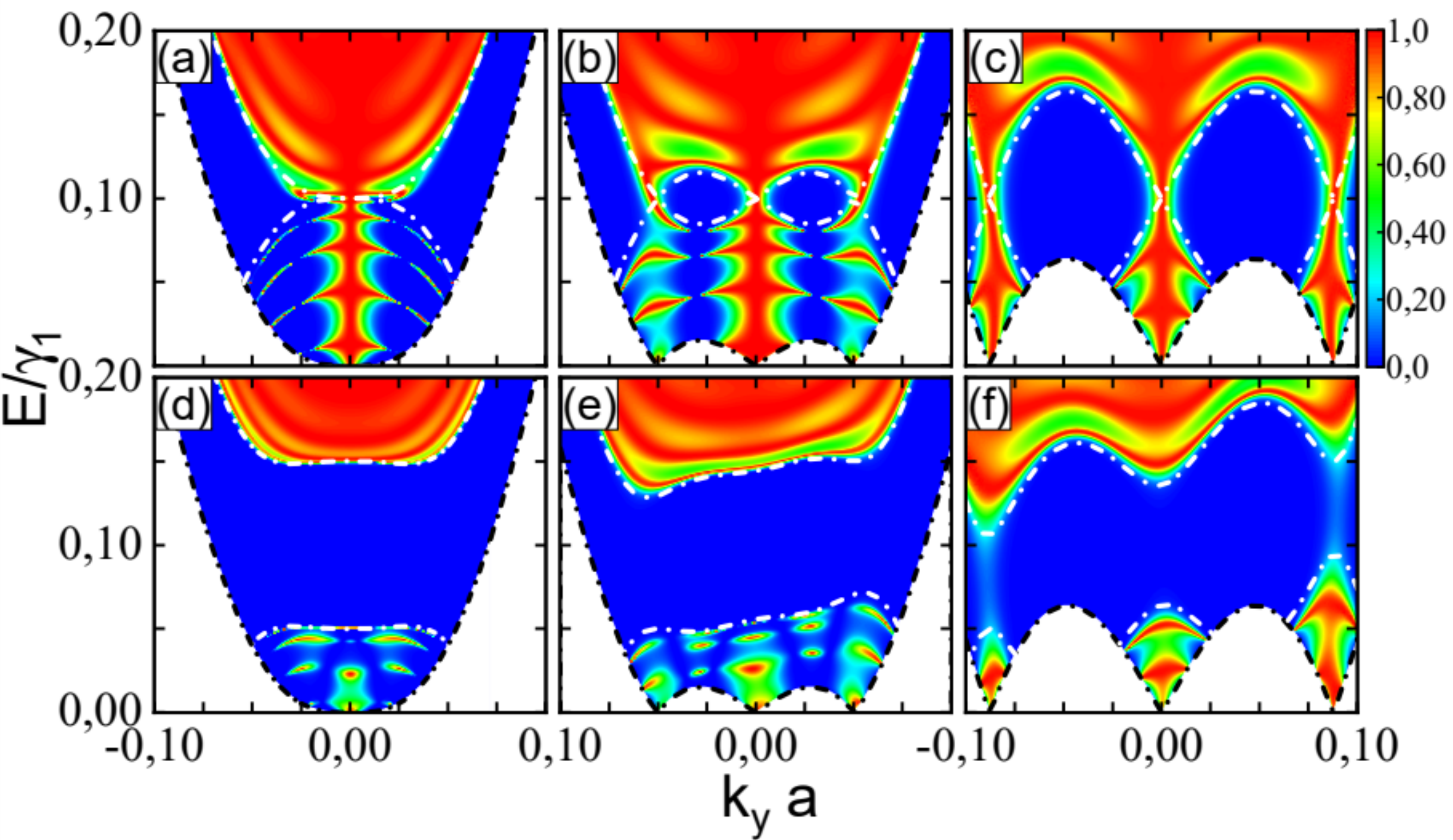}
\vspace{-0.5cm}
\caption{(Color online) 
        Transmission of ABC-TLG in the parallel magnetic field $ B $ as function of the energy and transverse wave vector through a $pnp$ junction of height $V_0=0.1\gamma_1$ and width $d=50$nm with (top row) zero  and (bottom row) finite inter-layer bias $\delta=0.05\gamma_1$. Here, we chose $B=0$T, $800$T and $1400$T in the left, middle, and right row, respectively.}\label{TransmissionEKYbarrier}
\end{figure}

The zero temperature conductivity can be calculated using the Landauer-B\"{u}ttiker formula\cite{Blanter336}. 
The transmission $T^{m}_{n}(E,k_y)$ is then used to calculate the conductivity and Fano factor of $pn$ and $pnp$ junctions. They are given by
\begin{align}
        &\label{eq24}
G(E)=g_{0}\frac{w}{2\pi}\int_{-\infty}^{+\infty}dk_{y}\sum^3_{m,n=1}T^{m}_{n}(E,k_y),
\\
&\label{eq24}
F(E)=\frac{\int_{-\infty}^{+\infty}dk_{y}\sum^3_{m,n=1}T^{m}_{n}(E,k_y)(1-T^{m}_{n}(E,k_y))}{\int_{-\infty}^{+\infty}dk_{y}\sum^3_{m,n=1}T^{m}_{n}(E,k_y)},
\end{align}
with $m=1, 2, 3$ refers to the propagation mode. $T_n^m$ represents the transmission probability of a particle incident from the mode $k_n$ (subscript $n$ in $T_n^m$) and transmitted to the mode $k_m$ (superscript $m$ in $T_n^m)$\cite{vanduppen195439,Jellal534}, while $w$ defines the length of the sample in the $y$-direction, and the conductivity unit $g_0=4e^2/h$, the factor $4$ refers  to the valley and spin degeneracy in graphene. 
It is worth noting that at low energy, $E<\gamma'_1$ ($\gamma'_1= 0.918 \gamma_{1}$\cite{vanduppen195439,Jellal534}), there is only one mode of propagation corresponding to the wave vector $k_1$, whereas at high energy, $E>\gamma'_1$, there are three modes corresponding to the wave vectors $k_1$, $k_2$.
%
 In the following section, we will show that at high energy, the effect of the parallel magnetic field on the transmission is less important than the one in the low energy case. For this reason, we will focus on this last case in our future analysis.

\section{RESULTS AND DISCUSSION}
\label{numerical results}

\begin{figure}[t!]
\vspace{0.cm}
\centering\graphicspath{{./Figures/}}
\includegraphics[width=2.7in]{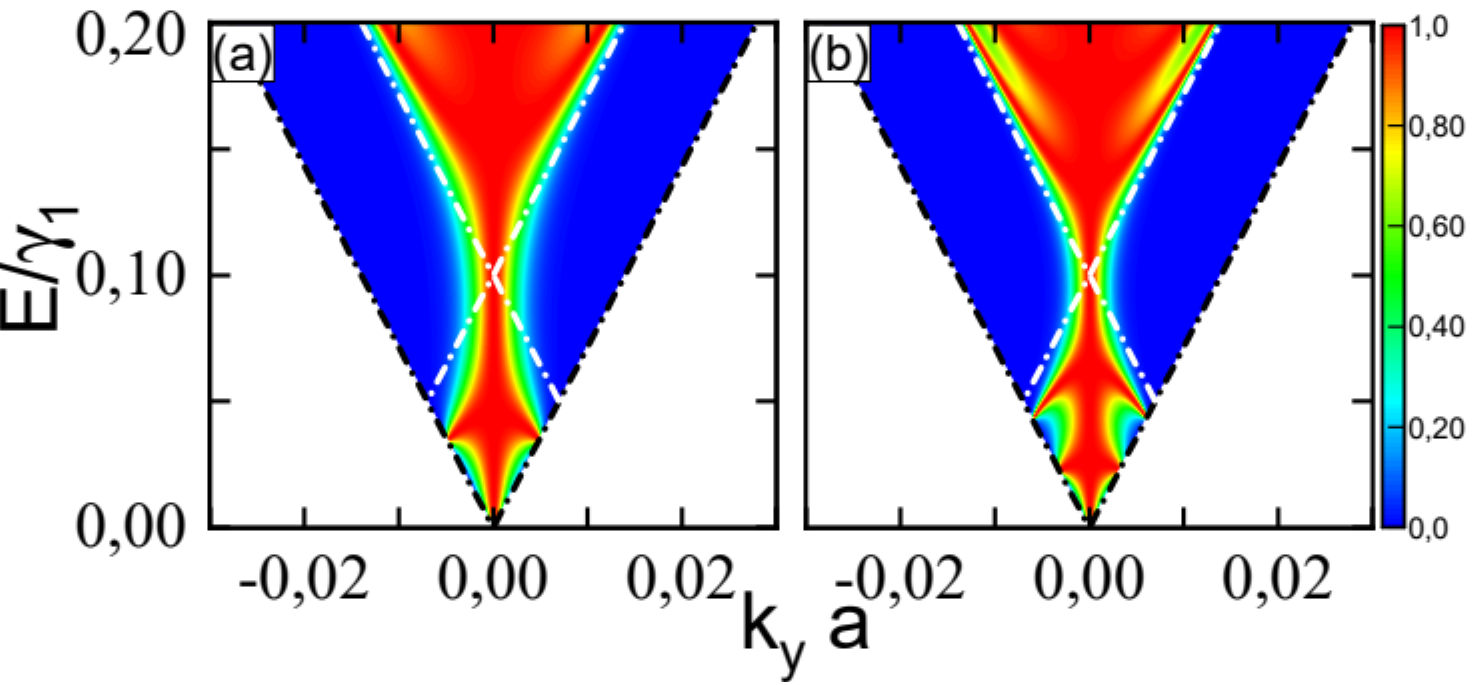}
\vspace{-0.5cm}
\caption{(Color online) 
        Transmission for SLG with $B=0$ as a function of energy and transverse wave vector through a $pnp$ junction of height $V_0=0.1\gamma_1$ and width (a): $d'=2d$   and (b): $d'=3d$  with $d=50$nm.}  \label{TransmissionSLGbarrier}
%
\end{figure}

Fig. \ref{TransmissionEKYbarrier} depicts the transmission of ABC-TLG through a $pnp$ junction with height $V_0=0.1\gamma_1$ and width $d=50$nm in a parallel magnetic field $B$ with inter-layer bias $\delta=0$ in (a,b,c) and $\delta=0.05\gamma_1$ in (d,e,f). 
%
It is worth noting that the results obtained in \cite{Kumar163102,Kumar222101,vanduppen195439,Jellal534} can be reproduced by tacking $B=0$.
In the absence of inter-layer bias and for a normal incidence ($k_y=0$) without ($B=0$T in (a)) and with ($B=800$T in (b) and $B=1400$T in (c)) magnetic field, the transmission is unit and becomes independent of energy, which is the same as in\cite{Kumar163102,vanduppen226101,vanduppen195439}.
This is a manifestation of Klein tunneling, which occurs for rhombohedrally stacked multilayers with an odd number of layers due to pseudospin conservation \cite{Kumar163102,vanduppen226101,vanduppen195439}.  Whatever the value of the energy, a new full transmission can occur at a non-normal incidence  $k_y=\pm\kappa$ where $\kappa=\pm0.050a$ for $B=800$T and $\kappa=\pm0.087a$ for $B=1400$T. 
In addition, the magnetic field splits the band energy into  three located at $k_y=0$ and at $k_y=\pm\kappa$ as depicted in Figs \ref{TransmissionEKYbarrier}(b,c), which correspond to Figs. \ref{EnergyBdelta}(b,c). 
At high magnetic fields, as shown in Fig. \ref{TransmissionEKYbarrier}(c), the transmission of an ABC-TLG through a $ pnp $ junction with height $V_0=0.1\gamma_1$ and width $d=50$nm behaves exactly like that of three separated SLG with $ B=0 $, except that the width is now $d'=2d=100$nm localized at $k_y=\pm\kappa$ for two SLG and $d'=3d=150$nm localized at $ky=0$ for one SLG. To clarify this statement, we plot in
Fig. \ref{TransmissionSLGbarrier}(a,b) 
the transmission  in SLG without $B$ via $pnp$ junction of height $V_0=0.1\gamma_1$ and width $d'=100,150 $nm\cite{Castro109,Katsnelson20,Katsnelson620}.
It is worth noting that for large enough magnetic fields of $  B = 600  $T\cite{Donck115423},  the AB-BLG system subjected to a parallel magnetic field is equivalent to two separated single-like systems, with Klein tunneling arising at $k_y=\pm\kappa$. 
\begin{figure}[t!]
\vspace{0.cm}
\centering\graphicspath{{./Figures/}}
\includegraphics[width=\linewidth]{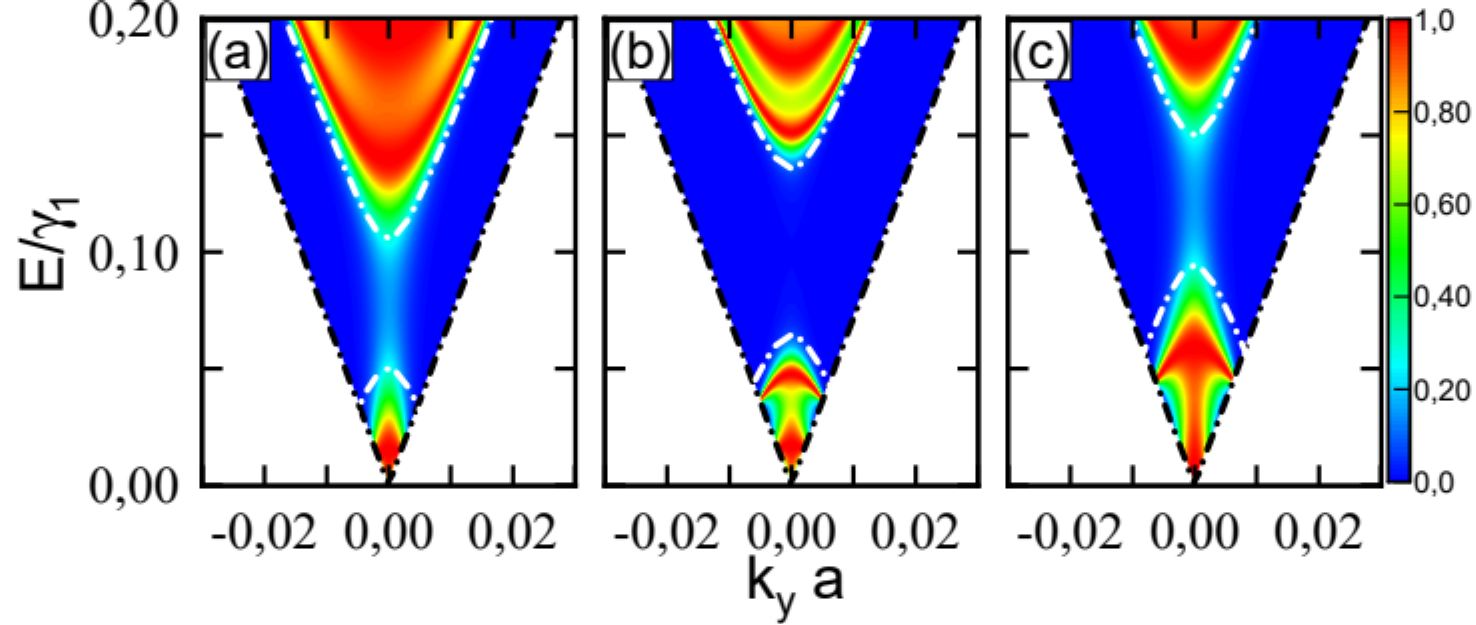}
\vspace{-0.5cm}
\caption{(Color online) Transmission for SLG with $B=0$ as a function of energy and transverse wave vector through a $pnp$ junction of height and inter-layer bias (a): $V_{0_1}=0.077\gamma_1$ and $\delta_{1}=0.027\gamma_1$, (b): $V_{0_2}=0.1\gamma_1$ and $\delta_{2}=0.035\gamma_1$, (c):  $V_{0_3}=0.122\gamma_1$ and $\delta_{3}=0.05\gamma_1$. The junctions width (a,c): $d'=2d$  and (b): $d'=3d$   with $d=50$nm.}\label{TransmissionSLGbarrierdelta}
\end{figure}

When $\delta=0.05\gamma_1$ is used without $B$, a direct gap in the transmission is created, as shown in Fig. \ref{TransmissionEKYbarrier}(d), whereas when $B=800$T is used, the energy spectrum becomes tilted, as shown in Fig. \ref{TransmissionEKYbarrier}(e). 
%
For $B=1400$T, we obtain three indirect gaps $E_{g_1}=0.055\gamma_1$, $E_{g_2}=0.071\gamma_1$ and $E_{g_3}=E_{g_1}$, which are clearly visible at $k_y=-\kappa$, $0$ and $+\kappa$  as depicted in Fig. \ref{TransmissionEKYbarrier}(f). This actually corresponds to the energy depicted in Fig. \ref{EnergyBdelta}(f). 
In addition, the transmission of ABC-TLG with $B\neq 0$ is split to a transmission of three SLG through a $pnp$ junction with the same previously chosen widths for $B=0$. For instance a large value $B=1400$T, we determine the height and inter-layer bias for each layer as follows:
 ($V_{0_1}=0.077\gamma_1$, $\delta_{1}=0.027\gamma_1$), ($V_{0_2}=V_0$, $\delta_{2}=0.035\gamma_1$) and ($V_{0_3}=0.122\gamma_1$, $\delta_{3}=\delta$) localized at   $k_y=-\kappa$, $0$ and $+\kappa$, respectively, with $\kappa=0.087a$.
%
These $V_{0_i}$ and $\delta_i$ values are determined by the minimum of the conduction band and the maximum of the valence band of ABC-TLG, and their formulas are shown in Fig. \ref{EnergyBdelta}.
 As a comparison,  we plot the transmission  for electrons in SLG through $pnp$ junction in Fig. \ref{TransmissionSLGbarrierdelta}(a,b,c) \cite{Zarenia085451, Azarova118} with the same previous values of $d'$, $V_{0_i}$ and $\delta_i$. 
%
%
\begin{figure}[t!]
\vspace{0.cm}
\centering\graphicspath{{./Figures/}}
\includegraphics[width=\linewidth]{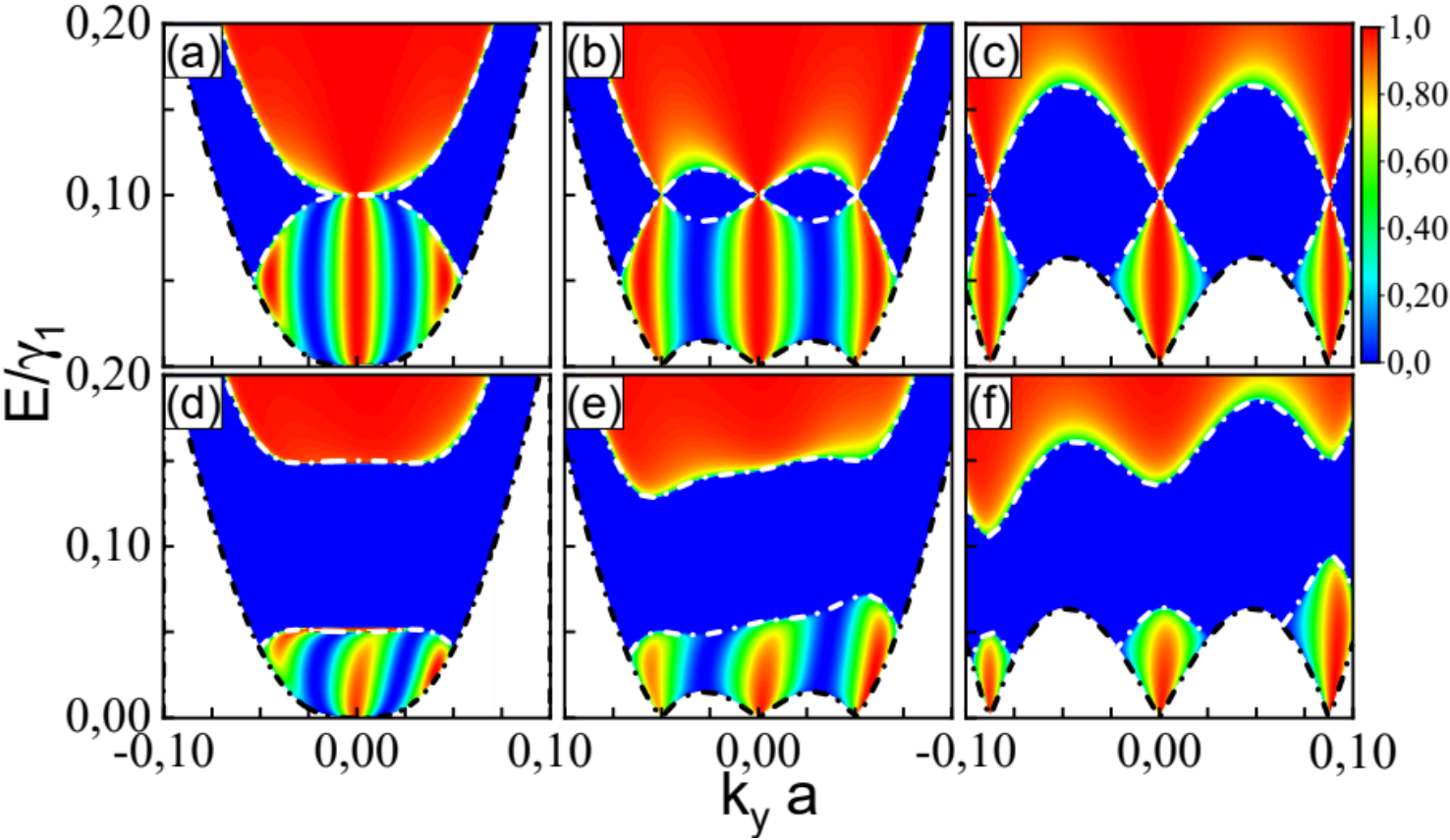}
\vspace{-0.5cm}
\caption{(Color online) The same as Fig. \ref{TransmissionEKYbarrier} but now through a $pn$ junction.}\label{TransmissionEKYstep}
\end{figure}
\begin{figure}[t!]
\vspace{0.cm}
\centering\graphicspath{{./Figures/}}
\includegraphics[width=1.7in]{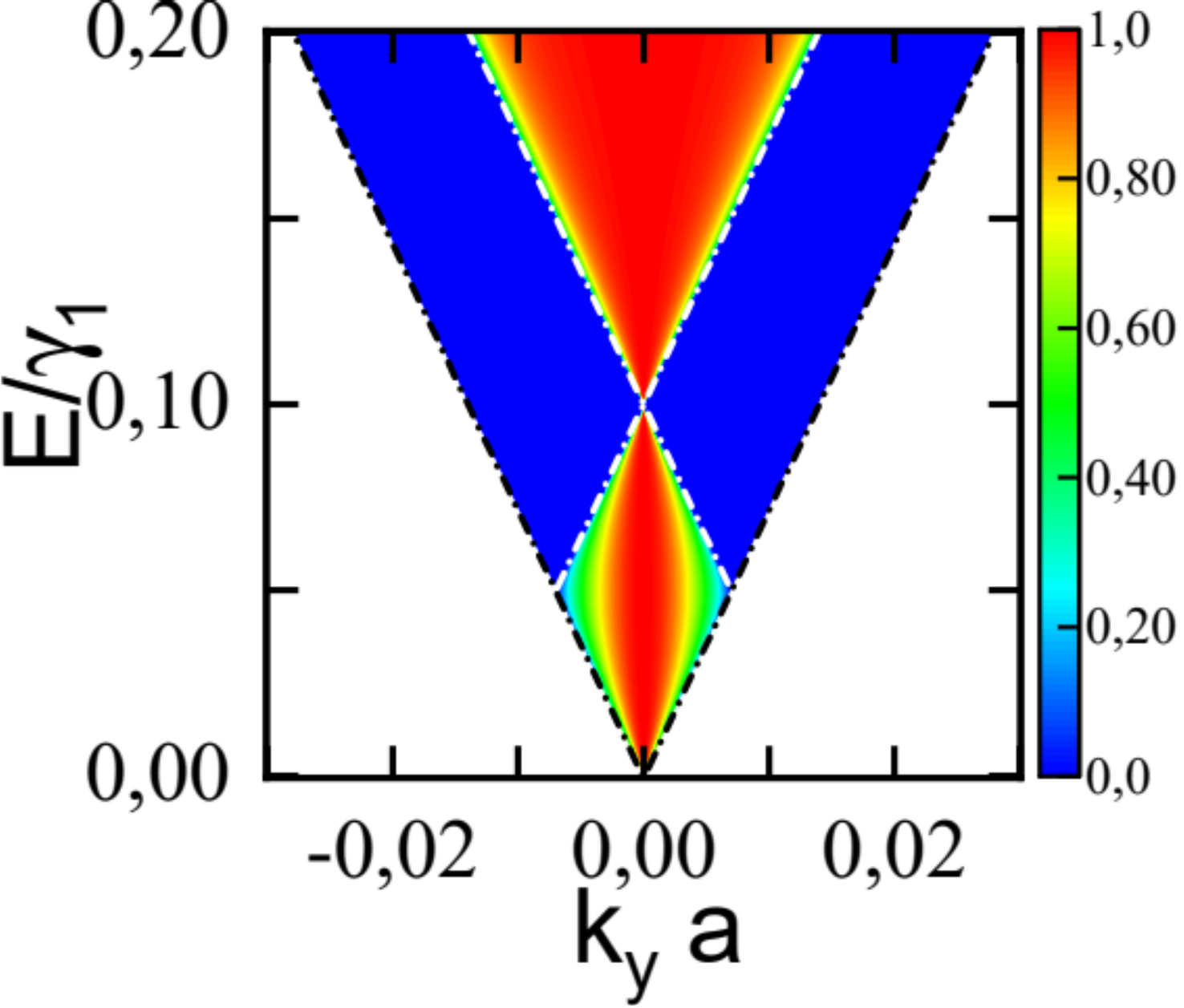}
\vspace{-0.5cm}
\caption{(Color online) The same as Fig. \ref{TransmissionSLGbarrier} but now through a $pn$ junction.}\label{TransmissionSLGstep}
\end{figure}
\begin{figure}[t!]
\vspace{0.cm}
\centering\graphicspath{{./Figures/}}
\includegraphics[width=\linewidth]{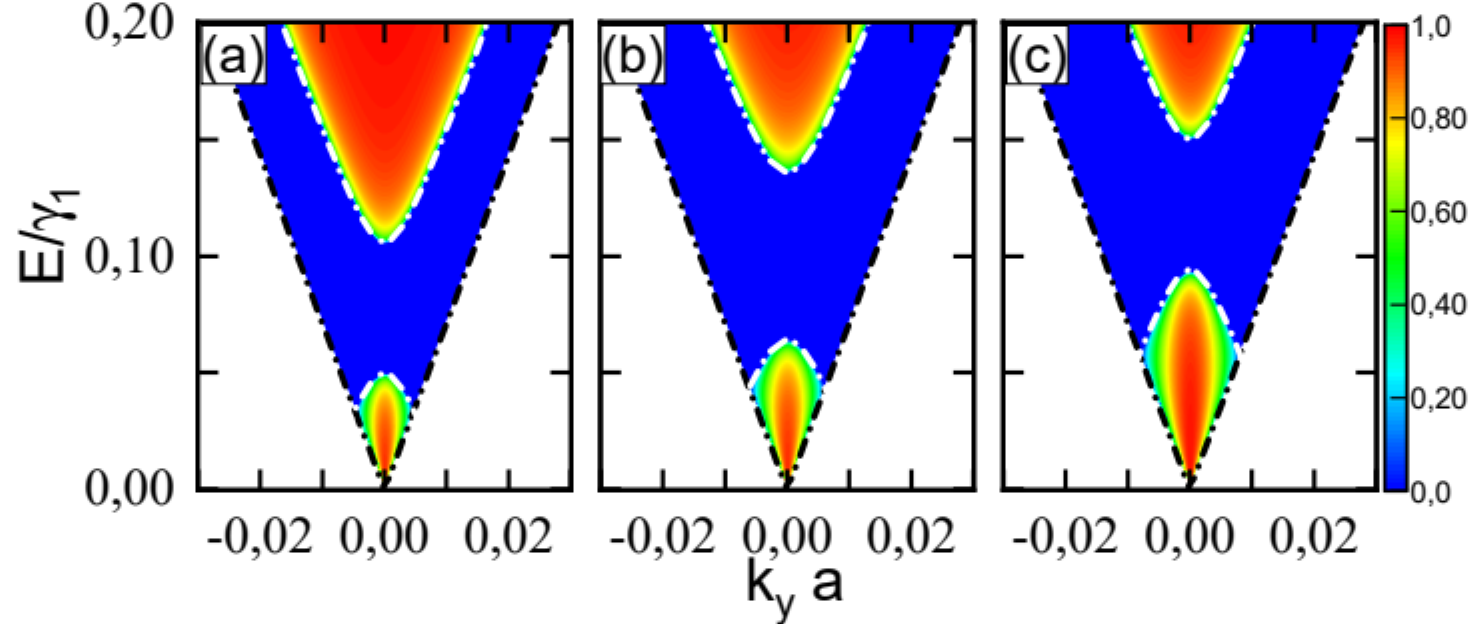}
\vspace{-0.5cm}
\caption{(Color online) The same as Fig. \ref{TransmissionSLGbarrierdelta} but now through a $pn$ junction.}\label{TransmissionSLGstepdelta}
\end{figure}

Fig. \ref{TransmissionEKYstep} depicts ABC-TLG transmission through a $pn$ junction of height $V_0=0.1\gamma_1$ with $\delta=0$ in (a,b,c) and $\delta=0.05\gamma_1$ in (d,e,f). 
The previously discussed equivalence of ABC-TLG to three SLG at high magnetic field with and without inter-layer bias holds true in the $pn$ junction with the same $V_{0_i}$ and $\delta_i$ values as in the $pnp$ junction. 
 To be more specific, transmission in ABC-TLG at a high parallel magnetic field through a $ pn $ junction is equivalent to transmission in three-SLG localized at $k_y=-\kappa $, $0$, and $+\kappa $ as in  $pnp$ junction. 
%
%
Figs. \ref{TransmissionSLGstep} and \ref{TransmissionSLGstepdelta}, for example, show transmission in MLG with $B=0$ through a $pn$ junction.  In Fig. \ref{TransmissionSLGstep}, 
the junction height and inter-layer bias are $V_0=0.1\gamma_1$ and $\delta=0$, whereas in Fig. \ref{TransmissionSLGstepdelta} we have (a) $V_{0_1}=0.077\gamma_1$ and $\delta_{1}=0.027\gamma_1$, (b) $V_{0_2}=0.1\gamma_1$ and $\delta_{2}=0.035\gamma_1$, (c)  $V_{0_3}=0.122\gamma_1$ and $\delta_{3}=0.05\gamma_1$.

\begin{figure}[tbh]
\centering
\includegraphics[width=3.45in]{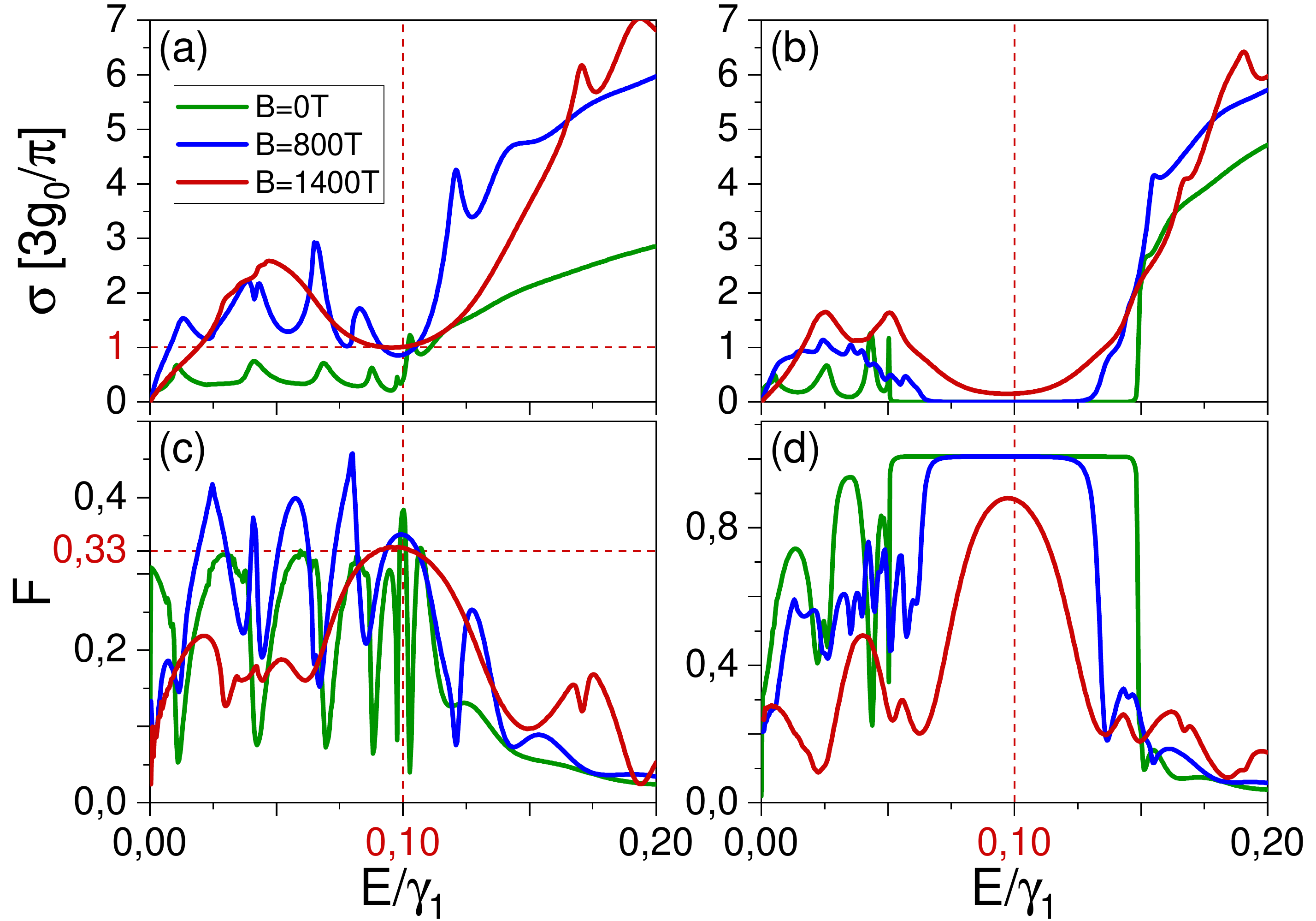}
\caption{(Color online) 
        The energy dependence of conductivity (top row) and Fano factor (bottom row) for the ABC-TLG in a parallel magnetic field and through a $pnp$ junction of height $V_0=0.1\gamma_1$ and width $d=50$nm.  $\delta=0$ (left panel), $\delta=0.05\gamma_1$ (right panel), $B= 0T$ (green line),  $800T$ (blue line) and $1400T$ (red line) were chosen. The horizontal line in (a) corresponds to $\sigma=3g_0/\pi$, whereas the horizontal line in (c) corresponds to $F=1/3$. 
}\label{ConduFanopnp}
\end{figure}
\begin{figure}[tbh]
\centering
\includegraphics[width=3.45in]{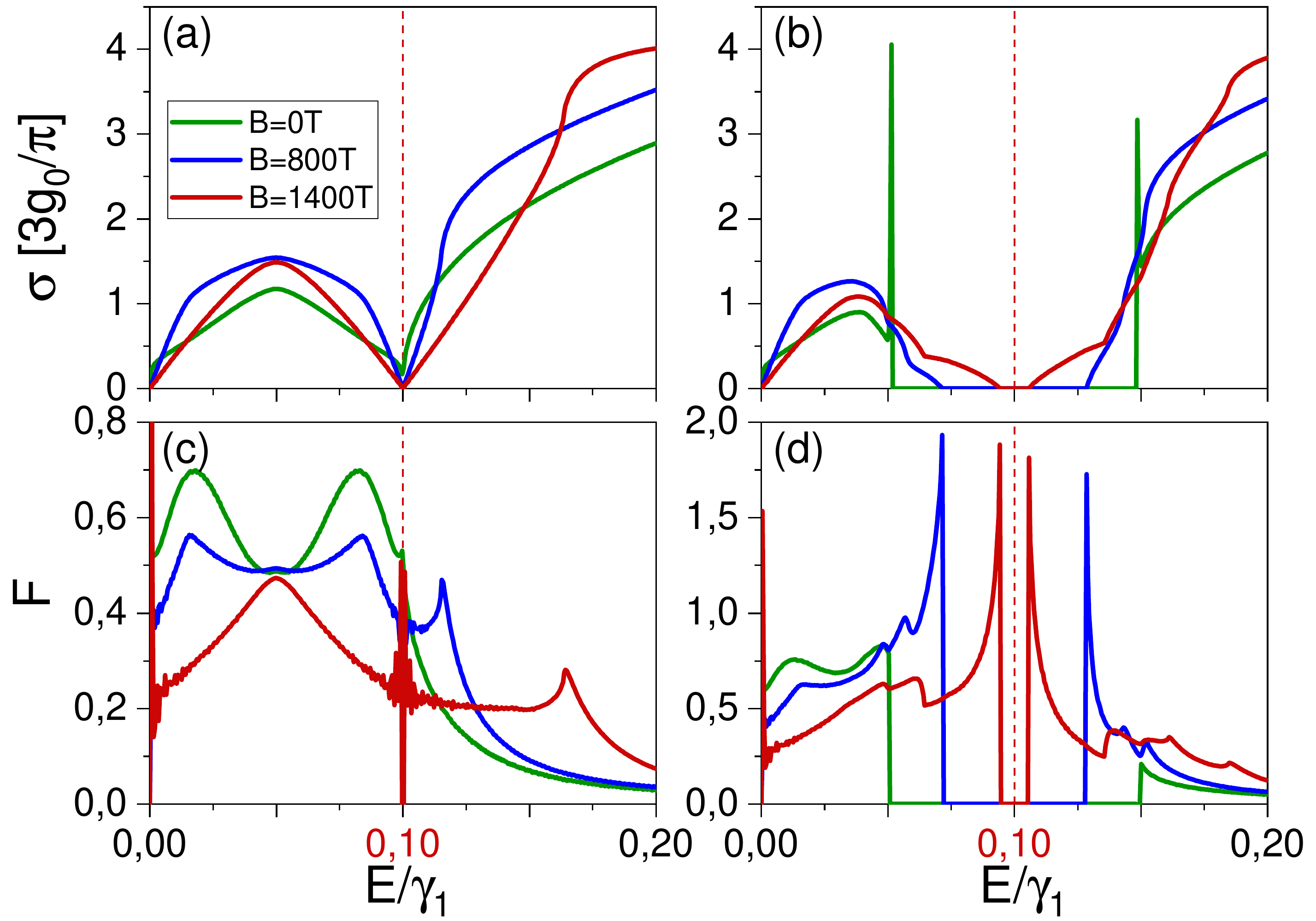}
\caption{(Color online) The same as Fig. \ref{ConduFanopnp} but now through a $pn$ junction.}\label{ConduFanopn}
\end{figure}
The conductivity $\sigma=G\ d/w$  and Fano factor in ABC-TLG through $pnp$ and $pn$ junctions of height $V_0=0.1\gamma_1$ and width $d=50$nm with and without $\delta$ as a function of the Fermi energy for different values of the applied magnetic field $B$ are shown in Figs. \ref{ConduFanopnp} and \ref{ConduFanopn}. 
%
When  $B$ is less than $1400 $T, the conductivity and  Fano factor show several sharp peaks in the $ pnp $ junction, while for $B\geqslant1400$T, both quantities  behave like those in SLG\cite{Beenakker246802,Danneau196802} as shown in Figs. \ref {ConduFanopnp}(a,c) for $\delta=0$ and in Figs. \ref{ConduFanopnp}(b,d)  for $\delta=0.05\gamma_1$. 
%
Furthermore, it has been shown that, similarly to without a magnetic field, the ABC-TLG scattered by a single barrier with a parallel magnetic field also has a minimum conductivity associated with a maximum Fano factor. The conductivity minimum $g_0/\pi$ and Fano factor $1/3$ observed in SLG\cite{Beenakker246802,Danneau196802} through the $pnp$ junction are remarkably reproduced at a high magnetic field of $B=1400$T applied to ABC-TLG, with $\sigma_{\text{TLG}}=3\sigma_{\text{SLG}}$.   
%
The induced gap in transmission, as shown in Fig. \ref{TransmissionEKYbarrier}(f) at high magnetic field with inter-layer bias, has resulted in the formation of a conductivity other than zero at $V_0=0.1\gamma_1$, as illustrated in Fig. \ref{ConduFanopnp}(b). 
%

In contrast, in the $ pn $ junction without inter-layer bias, as shown in Fig. \ref{ConduFanopn}(a), the conductivity goes to zero at $E=0$ and $E=V_0$, which is the potential step height, similarly to SLG\cite{Beenakker246802,Danneau196802}, AB-BLG and ABC-TLG\cite{vanduppen195439,Jellal534} with $B=0$. 
%
This is in agreement with Figs. \ref{TransmissionEKYstep}(a,b,c), which show that the effect of the parallel magnetic field is strongest at $E=V_0$, which leads to a suppression of electron transmission for all values of $k_y$. 
Furthermore, in the energy range of $0<E<V_0$, the conductivity of ABC-TLG and SLG is maximum at the mid-barrier height of $E=V_0/2$ in the presence of a parallel magnetic field. 
We can clearly see in Fig. \ref{ConduFanopn}(b) that the conductivity of a biased ABC-TLG in a parallel magnetic field increases in the energy range of $V_0-\delta$ to $V_0+\delta$, and some of the main peaks vanish as a result of a high parallel magnetic field and the induced gap in the transmission due to inter-layer bias $\delta$. 
%
%
For a $pn$ junction, the correspondence between the minimum conductivity and the maximum Fano factor will remain valid, as shown in Figs. \ref{ConduFanopn}(a,b) and in Figs. \ref{ConduFanopn}(c,d), respectively.

\section{conclusions}
\label{conclusions}

We have studied the Klein tunneling effect at low energy in ABC-TLG through a $pn$ and $pnp$ junctions with parallel magnetic field using the transfer matrix  method. As a result, As a result, we discovered that ABC-TLG is  $100\%$ transparent without inter-layer bias and at $k_y=0$ and $k_y=\pm\kappa$ with $\kappa=\pm0.050a$ for $B=800$T and $\kappa=\pm0.087a$ for $B=1400$T.
When an inter-layer bias is used, a gap is created, making transmission impossible at $k_y =0$ for $pn$ and $pnp$ junctions, and at $k_y=\pm\kappa$ for $pn$ junctions only. Transmission is possible at $k_y=\pm\kappa$, unlike the $pnp$ junction. 

%
%

Subsequently, we discovered a very pronounced transition from a trilayer system to three separated monolayer-like systems at $k_y=0$ and $k_y=\pm \kappa$ regardless of the energy values at a high magnetic field of $B=1400$T. 
The usual minimum  conductivity $g_0/\pi$ and maximum value $1/3$ 
of Fano factor for SLG are reproduced when the ABC-TLG is subjected to a large parallel magnetic field $B=1400$T, with $\sigma_{\text{TLG}}=3\sigma_{\text{SLG}}$.
The results presented here are potentially exploitable for paving the way for electrical control of quantum transport in ABC-TLG-based electronic devices.

\section{Acknowledgments}
The generous support provided by the Saudi Center for Theoretical
Physics (SCTP) is highly appreciated by all authors.

\end{document}